\documentclass{nature}
\usepackage[hidelinks]{hyperref}
\usepackage{graphicx}
\usepackage{color}
\usepackage{amsmath}
\usepackage{amssymb}
\usepackage{rotating}
\usepackage{longtable}
\usepackage{multibib}
\usepackage{booktabs}
\newcites{Supp}{~}
\usepackage{float}
\usepackage{upgreek}
\usepackage{caption}
\title{Detection of titanium oxide in the atmosphere of a hot Jupiter}

\author{Elyar Sedaghati$^{1,2,3}$, Henri M. J. Boffin$^1$, Ryan J. MacDonald$^4$, Siddharth Gandhi$^4$, Nikku Madhusudhan$^4$, Neale P. Gibson$^5$, Mahmoudreza Oshagh$^{6,7}$, Antonio Claret$^8$, \& Heike Rauer$^{2,3}$}

\begin{document}

\maketitle
\begin{affiliations}
 \item European Southern Observatory, Alonso de C\'ordova 3107, Santiago, Chile
 \item Deutsches Zentrum f\"ur Luft- und Raumfahrt, Rutherfordstr. 2, 12489 Berlin, Germany
 \item Zentrum f\"ur Astronomie und Astrophysik, TU Berlin, Hardenbergstr. 36, 10623 Berlin, Germany
 \item Institute of Astronomy, University of  Cambridge, Madingley Road, CB3 0HA Cambridge, UK
 \item Astrophysics Research Centre, School of Mathematics and Physics, Queens University Belfast, BT7 1NN Belfast, UK
 \item Institut f\"ur Astrophysik, Georg-August Universit\"at G\"ottingen, Friedrich-Hund-Platz 1, 37077 G\"ottingen, Germany
 \item Instituto de Astrof\' isica e Ci\^encias do Espa\c{c}o, Universidade do Porto, CAUP, Rua das Estrelas, 4150-762 Porto, Portugal
 \item Instituto de Astrof\'isica de Andaluc\'ia, CSIC, Apartado 3004, 18080 Granada, Spain
\end{affiliations}

\begin{abstract}
As an exoplanet transits its host star, some of the light from the star is absorbed by the atoms and molecules in the planet’s atmosphere, causing the planet to seem bigger; plotting the planet’s observed size as a function of the wavelength of the light produces a transmission spectrum\cite{Seager2000}. Measuring the tiny variations in the transmission spectrum, together with atmospheric modelling, then gives clues to the properties of the exoplanet’s atmosphere. Chemical species composed of light elements—such as hydrogen, oxygen, carbon, sodium and potassium—have in this way been detected in the atmospheres of several hot giant exoplanets\cite{Deming2013,Kreidberg2014,Madhusudhan2016,Sing2016}, but molecules composed of heavier elements have thus far proved elusive. Nonetheless, it has been predicted that metal oxides such as titanium oxide (TiO) and vanadium oxide occur in the observable regions of the very hottest exoplanetary atmospheres, causing thermal inversions on the dayside\cite{Hubeny2003,Fortney2008}. Here we report the detection of TiO in the atmosphere of the hot-Jupiter planet WASP-19b. Our combined spectrum, with its wide spectral coverage, reveals the presence of TiO (to a confidence level of 7.7$\upsigma$), a strongly scattering haze (7.4$\upsigma$) and sodium (3.4$\upsigma$), and confirms the presence of water (7.9$\upsigma$) in the atmosphere\cite{Huitson2013,Sing2016}.
\end{abstract}

Hot Jupiters are gas-giant exoplanets with sizes like that of Jupiter but much shorter orbital periods. WASP-19b is the shortest-period hot Jupiter to be discovered so far\cite{Hebb2009}, and has an excessively bloated radius, owing to the extreme radiation that it receives from its host star; as a result of this radiation, the planet’s effective temperature is more than 2,000 K (obtained via secondary-eclipse measurements\cite{Wong2016}). It is thought that high atmospheric temperatures imply the presence of metal oxides such as TiO, but despite extensive searches\cite{Haynes2015,Evans2016} a definitive detection of metal oxides in exoplanetary atmospheres has proved elusive.

We observed three transits of WASP-19b with the 8.2-metre Unit Telescope 1 (UT1) of the European Southern Observatory’s Very
Large Telescope (VLT), using the low-resolution FORS2 spectrograph. By using three of FORS2’s grisms—600B (blue), 600RI (green) and 600z (red), thereby covering the entire visible-wavelength domain (0.43–1.04 $\mu$m)—together with the multi-object spectroscopy configuration, we were able to obtain relatively high-resolution, precise, broadband transmission spectra. Such results were made possible through optimized observing strategies\cite{Boffin2016} and careful design of the observing mask used for the multi-object observations: this has slits about 30$^{\prime\prime}$ wide, which minimized differential losses owing to variations in telescope guiding and seeing conditions. The observations presented here were made between 11 November 2014 and 29 February 2016.

For each set of observations, we obtained a series of spectra for the main target (WASP-19), as well as for several comparison stars.
After standard data-reduction steps, we integrated those spectra for the largest common wavelength domain and 10-nm bins, to produce
the `white' and `spectrophotometric' light curves, respectively. To correct for the imprint of telluric variations on the detected light, we
divided the light curve of the target star by that of the one comparison star providing the best result. This corrected light curve was chosen through a detailed, statistical analysis of the differential light curves of all comparison stars, as well as all their possible combinations\cite{Sedaghati2016,Sedaghati2017}.

\begin{figure}
\includegraphics[width=\textwidth]{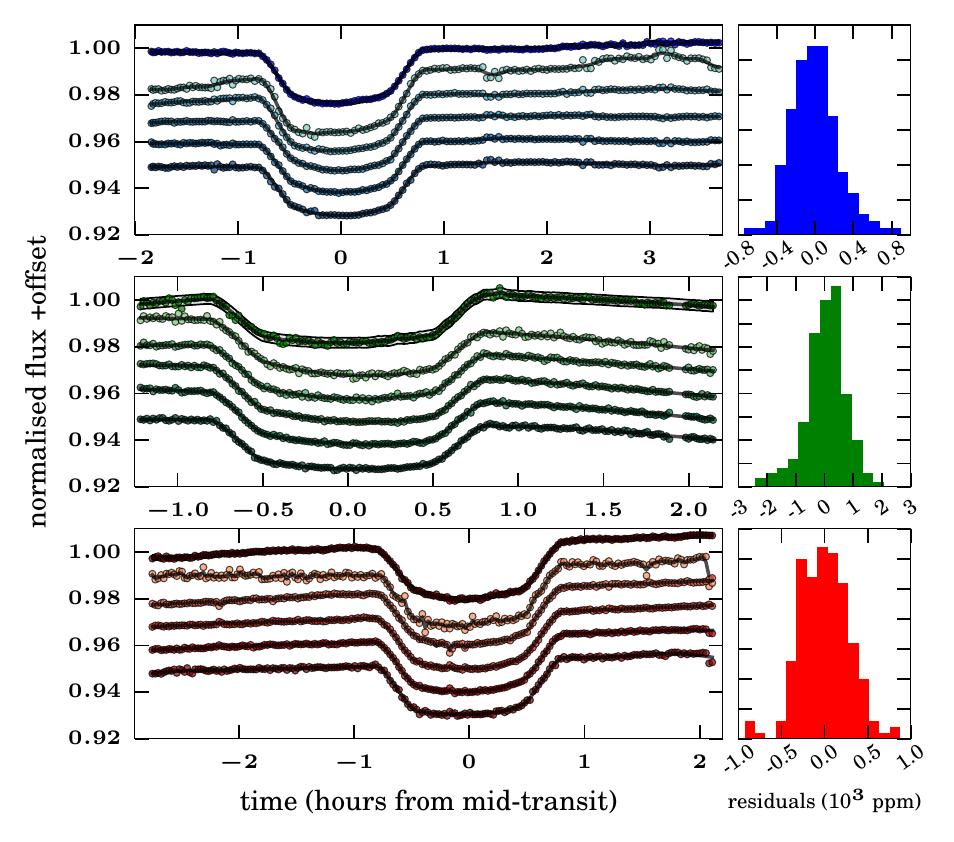}
\caption{\textbf{Light curves and models.}  Broadband and spectrophotometric light curves of WASP-19b from our three transit observation campaigns. The colour of each row corresponds to the data set that it represents, obtained using the 600B (blue), 600RI (green) and 600z (red) grisms. In the left panels, for each set, the top plot is the broadband transit light curve, used for determining wavelength-independent parameters, and the others are demonstrative spectral light curves produced using 50-nm integration bins. The grey-shaded regions in the broadband plots highlight the 3$\upsigma$ confidence level of the Gaussian process model. The panels on the
right show the distribution of the residuals of the analytical models fit to the broadband light curves.}
\label{fig:LCs main}
\end{figure}

We initially fitted the broadband light curves by using an analytical solution for a transit\cite{Mandel2002}, as well as a Gaussian process model for the estimation of correlated noise in the data\cite{Gibson2012}. This approach provides a model-independent stochastic method for including the systematic component of noise into our model. We derived the maximum posterior probability distributions for the fitted parameters by solving the Bayesian relation and running multiple, very long Monte Carlo Markov Chain (MCMC) simulations of our multivariate model\cite{Cameron2007}. The Gaussian process systematic model is trained by using carefully selected inputs, which are those physical variants that play a role in introducing correlated noise to the time-series data\cite{Gibson2012HD189}. This method gives us a robust estimate of the uncertainties in determining the planetary physical and orbital parameters, by fully accounting for the contribution of systematic noise\cite{Gibson2012}. Once the optimal broadband solutions are found for each data set, we use the wavelength-independent parameter values as strict, informative priors in our Bayesian analysis of the spectral light curves, and solve for variations of the planetary radius as a function of wavelength bin. Broadband light curves for the three sets of observations, exemplar spectral light curves and the fitted systematic models are shown in Fig. \ref{fig:LCs main}.

\begin{figure}
\includegraphics[width=\textwidth]{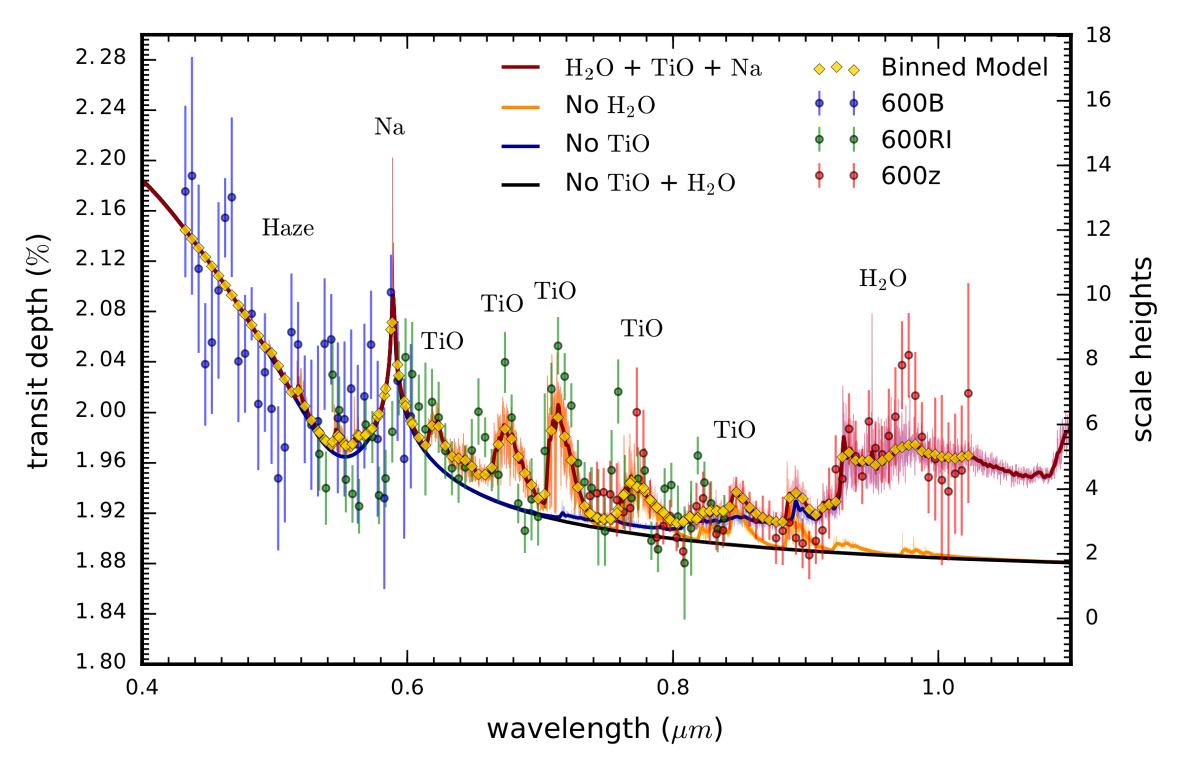}
\caption{\textbf{WASP-19b transmission spectrum}. Blue, green and red data points correspond to observations made using the 600B, 600RI and 600z grisms, respectively. The associated 1$\upsigma$ error bars were derived from posterior probability distributions of the planetary radius parameter in a joint analysis of the MCMC chains, where the mean of each distribution is plotted as the solution. The overall best-fitting spectrum--which includes opacity resulting from the presence of H$_2$O, TiO, Na, and a global haze--is shown as a red curve and yellow points at a representative resolution of about 3,000. Other curves (orange, blue and black) represent models with specific species removed. We applied a corrective wavelength shift between the model and the data of 73.6 \AA.}
\label{fig:transmission_spectrum}
\end{figure}

Figure \ref{fig:transmission_spectrum} shows the wavelength-dependent radius variations—the transmission spectrum—where the combination of results from the three observing campaigns constructs a broadband spectrum with high precision. It is this precision that facilitates the detection of features originating from the exoplanetary atmosphere. Such signals are imprinted upon the observed spectrum owing to the discrete absorption of the host star’s light by gases as the light traverses the day–night boundary of the exoplanetary atmosphere\cite{Seager2000}. Interpreting this transmission spectrum requires fitting theoretical atmospheric models to the data, a process that involves exploring a wide range of the atmospheric parameter space, described shortly. However, before performing this task, an important factor to consider is the role that the host star's activity plays in introducing spurious signals into our final spectrum\cite{Oshagh2013}. Especially important is the presence of stellar active regions, be it spots (dark) or faculae (bright). If the transiting planet happens to traverse one of these regions from the point of a view of an observer on Earth, then the light curve shows an anomaly in the form of a small bump or a dip. For instance, we see an evidence for such a spot-crossing event (a bump) in the blue light curve, where the anomaly cannot be attributed to changes in observing conditions. To circumvent this, we include an additional complexity in our analytical model for a circular-shaped spot on the stellar surface with a temperature below that of the surrounding stellar photosphere. This ensures that our results are not biased by the spot-crossing event. We also account for the possible presence of active regions on parts of the stellar disk that are not covered by the transit chord, better known as unocculted spots or faculae. Their presence introduces a bias in determining the out-of-transit baseline flux, which in turn leads to errors in the inferred value of the transit depth and hence to errors in the estimated planetary radius. This is particularly important for transmission spectroscopy, as the effect is chromatic\cite{Oshagh2014}. Our analysis reveals that the enhanced planetary radius seen towards the ultraviolet end of our spectrum can be attributed only partially to the possibility of unocculted spots, even when assuming a large spot-filling factor and high temperature differences.

We infer the atmospheric properties of WASP-19b at the day–night boundary, probed by our transmission spectrum, by using the atmospheric retrieval algorithm POSEIDON\cite{MacDonald2017}. This algorithm explores many millions of transmission spectra--spanning a wide range of chemical compositions, temperatures, and cloud/haze properties--to identify the range of atmospheres that is consistent with the observations. In addition to standard absorbers expected in hot Jupiter atmospheres\cite{Madhusudhan2016} (H$_2$, He, Na, K and H$_2$O), we consider a wide range of metal oxides and hydrides--TiO, VO, AlO, TiH, NaH, MgH, CrH, CaH, ScH and FeH. Our models span the continuum from clear to cloudy atmospheres, both with and without scattering hazes, and include two-dimensional models with inhomogeneous cloud coverage.

\begin{figure}
\centering
\includegraphics[width=0.6\textwidth]{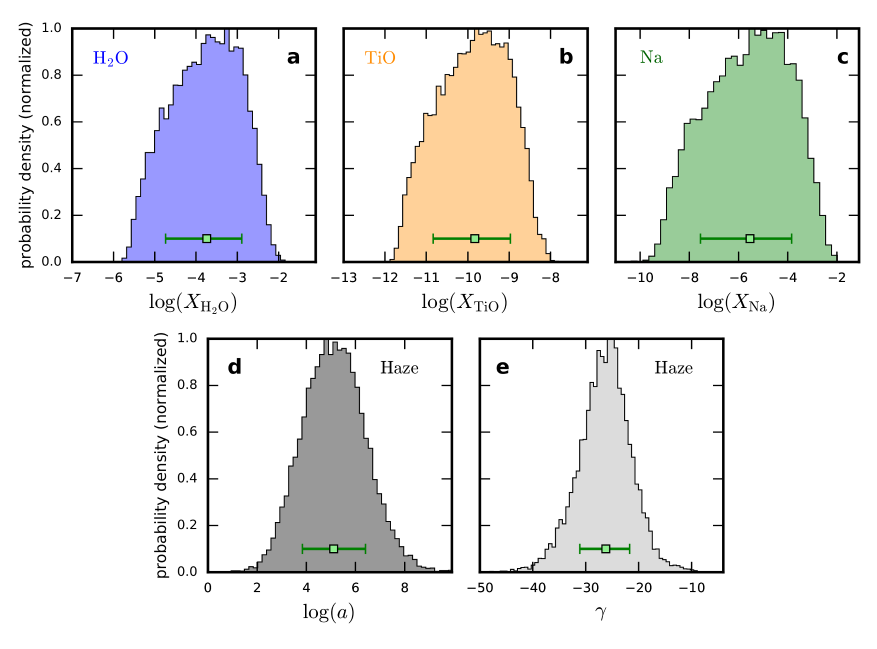}
\caption{\textbf{Constituents detected in WASP-19b's atmosphere.} \textbf{a--c}, Histograms showing the marginalized posterior probability densities of the H$_2$O (\textbf{a}), TiO (\textbf{b}) and Na (\textbf{c}) volume mixing ratios at WASP-19b's day-night terminator. \textbf{d}, \textbf{e}, Histograms showing the marginalized posterior probability densities of the haze Rayleigh enhancement factor, \textit{a}, and scattering slope\cite{LecavelierDesEtangs2008}, $\Upsilon$, as defined by the power law $ \sigma(\lambda) = a \, \sigma_{0} (\lambda/\lambda_{0})^\Upsilon $, where $\lambda_0$ is a reference wavelength (350 nm) and $\sigma_0$ is the H$_2$-Rayleigh scattering cross-section at the reference wavelength (5.31 $\times$ 10$^{-31}$ m$^2$).}
\label{fig:detection_histograms}
\end{figure}

We report detections of H$_2$O (confidence limit 7.9$\upsigma$), TiO (7.7$\upsigma$) and Na (3.4$\upsigma$), as well as a strongly scattering haze (7.4$\upsigma$) that envelopes the planet. Constraints on the volume mixing ratios and haze properties are presented in Fig. \ref{fig:detection_histograms}. The derived abundances of H$_2$O (18--1,300 p.p.m.) and Na (0.028--140 p.p.m.) span a wide range, at one end being consistent with expectations of an atmosphere of solar composition, and at the other being substantially sub-sola \cite{Madhusudhan2012}; the TiO abundances (the 1σ range is 0.015--1.1 p.p.b.) are sub-solar to a confidence limit of more than 5$\upsigma$ (the upper limit of which is 44 p.p.b.). Despite suggestions of substructure in the low-wavelength region of the spectrum (at around 0.5 $\mu$m or slightly higher), which could be explained by the presence of metal hydrides, our Bayesian model comparison leads to the conclusion that a haze optimally explains these observations, without the need to invoke other chemical species. The haze we detect is about 100,000 times stronger than Rayleigh scattering from H$_2$ alone, follows a power-law with an exponent of $-26^{+4}_{-5}$, and is consistent with 100\% coverage across the terminator. We do not detect an opaque cloud deck. Finally, the planet's atmospheric temperature in the line-of-sight at 1 mbar pressure is constrained to $2,350^{+168}_{-314}$ K.

Our detection of TiO in WASP-19b's atmosphere is consistent with expectations given the atmosphere’s high temperature (more than
2,000 K)\cite{Fortney2008}. Moreover, our spectroscopic detection of a refractory metal oxide in an exoplanetary atmosphere demonstrates the importance of visible-wavelength molecular opacity in transmission spectra. The strong visible opacity that results from the presence of TiO could have substantial effects on the temperature structure and circulation of the planet's atmosphere. If present in large enough quantities on the dayside, the TiO might cause a thermal inversion in the dayside atmosphere\cite{Hubeny2003,Fortney2008}, which could in the future be observable in high-precision thermal emission spectra obtained with the Hubble Space Telescope and James Webb Space Telescope. The increased opacity resulting from the presence of TiO could also lead to strong day-night temperature gradients in the atmosphere\cite{Showman2009}. On a more general note, our results demonstrate the usefulness of ground-based transmission spectroscopy for pursuing a detailed molecular analysis of exoplanetary atmospheres at optical wavelengths.

\begin{addendum}
\item[Online Content] Methods, along with any additional Extended Data display items and Source Data, are available in the online version of the paper; references unique to these sections appear only in the online paper.
\end{addendum}

\noindent \textbf{Received 27 April; accepted 6 July 2017.}
\vspace{1cm}

%\bibliography{WASP19}
%\bibliographystyle{naturemag}

\begin{addendum}
%\item[Supplementary Information] is linked to the online version of the paper at www.nature.com/nature.
\item This work is based on observations made with the FORS2 instrument on the European Southern Observatory (ESO)'s VLT. We thank staff astronomers J. Anderson and J. Smoker for performing some of the observations. E.S. acknowledges support from the ESO through the studentship programme. R.J.M. and S.G. acknowledge financial support from the UK Science and Technology Facilities Council (STFC) towards their doctoral programmes. M.O. acknowledges research funding from the Deutsche Forschungsgemeinschaft (DFG), grant OS 508/1-1, as well as support from the Funda\c c\~ ao para a Ci\^ encia e a Tecnologia (FCT) through national funds and from FEDER through COMPETE2020 from the following grants: UID/FIS/04434/2013 and POCI-01-0145-FEDER-007672; and PTDC/FIS-AST/1526/2014 and POCI-01-0145-FEDER-016886. We thank the Spanish Ministry of Education and Science (MEC; grants AYA2015-71718-R
and ESP2015-65712-C5-5-R) for support during the development of this work. We also thank the referees for their comments, which improved the manuscript.
 \item[Author Contributions] E.S. and H.M.J.B. led the scientific proposal, observational campaigns, data reduction and analysis up to the production of transmission spectra. R.J.M. conducted the atmospheric retrieval and S.G. generated the absorption cross-sections, both under the supervision of N.M., who planned and oversaw the atmospheric analyses and theoretical interpretation. N.P.G. wrote the python modules for the Gaussian process and the Monte Carlo Markov Chain analysis. M.O. analysed the impact of unocculted stellar active regions. A.C. calculated the theoretical limb-darkening coefficients for the specific bandpasses. H.R. provided feedback on the manuscript and is involved in the supervision of E.S. All authors contributed to writing the manuscript. 
 \item[Author Information] Reprints and permissions information is available at www.nature.com/reprints. The authors declare no competing financial interests. Readers are welcome to comment on the online version of the paper. Publisher's note: Springer Nature remains neutral with regard to jurisdictional claims in published maps and institutional affiliations. Correspondence and requests for materials should be addressed to E.S. (esedagha@eso.org).
 \item[Reviewer Information] \textit{Nature} thanks K. Heng and the other anonymous reviewer(s) for their contribution to the peer review of this work.
\end{addendum}

\newpage
\begin{methods}
%%%%%%%%%% Prefix a "S" to all equations, figures, tables and reset the counter %%%%%%%%%%
%\setcounter{equation}{0}
\setcounter{figure}{0}
\setcounter{table}{0}
\makeatletter
%\renewcommand{\theequation}{S\arabic{equation}}
%\renewcommand{\thefigure}{\arabic{figure}}
%\renewcommand{\bibnumfmt}[1]{[S#1]}
%\renewcommand{\citenumfont}[1]{S#1}
%%%%%%%%%% Prefix a "S" to all equations, figures, tables and reset the counter %%%%%%%%%%

\renewcommand{\figurename}{Extended Data Figure}
\renewcommand{\tablename}{Extended Data Table}

\noindent We here present, in greater detail, the observational, data-reduction and retrieval methods used in our analysis of WASP-19b's transmission spectrum.

\section*{Observations}

We have observed multiple transits of WASP-19b\cite{Hebb2009}, a hot Jupiter that is in an extremely tight orbit around its host star, WASP-19: it revolves around this magnitude-12.3 G8V star on a 0.789-day orbit, bringing it very close to its host's Roche limit. All observations were performed with ESO's FOcal Reducer and low dispersion Spectrograph (FORS2)\citeSupp{Appenzeller1998}, mounted at the Cassegrain focus of UT1 of the VLT. The data were obtained over a two-year period. Details of the three observing campaigns are given in Extended Data Table \ref{tab:obs}. The instrument has a $6.8^{\prime} \times 6.8^{\prime}$ field of view and comprises a mosaic of two 2k $\times$ 4k pixel detectors, which are either red optimized (MIT; green and red data) or blue optimized (E2V; blue data). The observations were prepared and mostly performed by members of our team.

For all sets of observations, the instrument’s atmospheric dispersion corrector (the LADC, which was upgraded in 2014\cite{Boffin2016}) was left in `park' position throughout the entire observing sequence, with its two prisms kept at a minimal separation of 30 mm. For the two most recent sets of observations, we chose to use the non-standard 200 kHz readout. This strategy allowed us to spend more time
on the target for a given cadence owing to the reduced readout time, subsequently increasing the signal-to-noise ratio of the spectra. For observations performed with the Mask eXchange Unit (MXU) mode, we created custom-designed masks with typically 30$^{\prime\prime}$-wide slitlets placed on top of the target star, and also on top of several reference stars and the sky in order to monitor variations in the background contamination level. As part of a routine calibration sequence, we also took bias, flat-field and arc images. The arc frames were taken with the same slit configuration as the scientific sequences, but with the width of the slitlets set to 1$^{\prime\prime}$ for increased resolution. Best effort was made in placing all of the observed stars in the middle of the detector set as much as possible, in order to ensure maximal wavelength overlap between the targets. An example of a mask created for observations in MXU mode is shown in Extended Data Fig. \ref{fig:observations}a. This is in keeping with the optimal observation strategy when using FORS2 for multi-object spectroscopy\cite{Boffin2016}. The observational conditions for all sets of observations were generally clear but not photometric. Some sporadic cirrus clouds affected the raw light curves in the `blue' campaign towards the end of the observing sequence; the influence of these clouds is mostly corrected for by differential photometry, although some residual effects remain.
These effects are however well modelled by the Gaussian process model. Finally, the results of the ‘green’ campaign are from the reanalysis of a data set that has been studied previously\citeSupp{Sedaghati2015}.

\section*{Data reduction}
For the purpose of data reduction, we wrote a specialized Pyraf pipeline\cite{Sedaghati2016,Sedaghati2017}, which included: first, overscan and bias-shape subtraction; second, spectroscopic flat-fielding; third, spectral extraction; and fourth, wavelength
calibration. For the extraction of one-dimensional spectra from the two-dimensional science frames, we implemented an optimal extraction algorithm\cite{Horne1986}. The size of the extraction box used for each star was varied, and chosen by analysing dispersion in the subsequent light curves (see Extended Data Fig. \ref{fig:observations}b for an example). Specifically, the width of the extraction box was selected once the gradient of the light curve dispersion relation to the extraction width was below 10$^{-4}$ pix$^{-1}$, for the frame with the largest point spread function in the spatial direction. This value was then used for all of the frames in the corresponding data set. In the example shown, 30 binned pixels were used for both the target and the comparison star. However, this extraction size varied from star to star, as well as from night to night, owing to variations in brightness and seeing conditions. Examples of such extracted spectra of the target and the comparison stars, for each data set, are shown in the top row of Extended Data Fig. \ref{fig:LCs}. To obtain the dispersion solutions we used a Chebyshev polynomial function fit of order 4, using the one-dimensional extracted arc frames, which provided results with a root mean square of 0.06 \AA~or better after deleting a few outlying features.

Once extracted, the series of the stellar spectra were integrated, initially within the largest common wavelength domain of all the targets. Their variations as a function of time result in the raw light curves. We then produced differential transit light curves by dividing the target light curve (that of WASP-19) by the light curves of all of the observed comparison stars, as well as all their possible combinations.
We chose the final comparison star as the one that produced the cleanest transit light curve (that is, the light curve with the lowest residuals relative to a transit model). Once chosen, we reintegrated the target and the chosen comparison star for their largest common wavelength overlap. Examples of these final raw and differential light curves are given in the middle and bottom rows of Extended Data
Fig. \ref{fig:LCs}, respectively, for the observing campaigns.

The spectrophotometric or narrow-band transit light curves were produced in an identical manner, but using integration bins of 10 nm. We chose this value through a statistical analysis of all three data sets, where the final decision was dictated by the least-precise set\cite{Sedaghati2016}. All of these light curves, used to produce the transmission spectrum, are shown in the left panels of Extended Data Figs \ref{fig:spec LCs - Blue}--\ref{fig:spec LCs - Red} for the blue, green and red data sets. We further corrected these light curves by
division through the residuals of the analytical model to the broadband light curve. This is known as common-mode systematic correction\citeSupp{Lendl2016,Nikolov2016,Gibson2017}, and it removes the wavelength-independent component of the systematic noise from the spectral light curves. The corrected light curves are shown in the right panels of Extended Data Figs \ref{fig:spec LCs - Blue}--\ref{fig:spec LCs - Red}, together with the Gaussian process models (described below).

\section*{Data analysis}
We modelled all transit light curves using a Gaussian process model, which provides a model-independent stochastic approach to including systematic signals\citeSupp{Rasmussen2006}. Their application to modelling exoplanet transit data and transmission spectroscopy has been demonstrated for the analysis of Hubble Space Telescope (HST)/Wide-Field Camera 3 (WFC3) data for probing the haze in the atmosphere of HD 189733b\cite{Gibson2012HD189}. We find maximum posterior solutions, $\mathcal{P}$, by optimizing
the Bayesian relation:

\begin{equation}\label{eq:log posterior}
\log \mathcal{P} \left( \theta, \phi | \boldsymbol{f},\boldsymbol{\mathsf{X}} \right) = \log \mathcal{L} \left( \boldsymbol{r} | \boldsymbol{\mathsf{X}}, \theta, \phi \right) + \log \mathrm{P} \left(\theta,\phi\right)
\end{equation} 

\noindent that is true up to a constant term, where $\theta$ and $\phi$ are collections of noise and transit model parameters respectively; $\boldsymbol{f}$ is the vector of flux measurements; $\boldsymbol{\mathsf{X}}$ is an N $\times$ K matrix (N being the number of observations and K the number of inputs needed to train the model); and $\boldsymbol{r}$ is the residual vector for the model, elements of which are calculated from flux measurements relative to an analytical transit model\cite{Mandel2002}. The two terms on the right-hand side are the likelihood function, $\mathcal{L}$, and prior probability, P, respectively. In order to introduce correlated noise to our covariance matrix, we write the likelihood function, $\mathcal{L}$, as a matrix equation:

\begin{equation}
\log \mathcal{L} \left( \boldsymbol{r} | \boldsymbol{\mathsf{X}}, \theta, \phi \right) = -\frac{1}{2} \left( \boldsymbol{r}^T \boldsymbol{\Sigma}^{-1} \boldsymbol{r} + \log |\boldsymbol{\Sigma}| + N\log 2\pi \right)
\end{equation}

\noindent where the superscript \textit{T} is the transpose and $\boldsymbol{\Sigma}$ represents the covariance matrix. Ideally one would like to know the full covariance matrix, and we simplify this process by `modelling' the elements of this N $\times$ N matrix via a covariance function, also more commonly known as the kernel ($k$). In this framework, the covariance matrix is written as $\Sigma_{ij} = k(x_i,x_j,\theta) + \delta_{ij}\sigma^2$, $x$ being an input of the kernel, $\delta$ the Kronecker delta and $\sigma^2$ the variance term. The last term ensures the addition of Poisson or white noise to the diagonal of the covariance matrix. Our choice for the kernel is the squared exponential (SE), which for a multi-dimensional parameter space (K; defined above), in its additive form, is written as:

\begin{equation}
k_{SE}(x_i,x_j,\theta) = \zeta \exp \left[ -\sum_{\alpha=1}^K \eta_\alpha \left( x_{\alpha,i}-x_{\alpha,j}\right)^2\right]
\end{equation}

We chose this form of the kernel as it is the \textit{de facto} default form for Gaussian processes, being infinitely differentiable and easy to integrate against most functions. In this definition, $\zeta$ is the maximum covariance and $\eta_\alpha$ are the inverse scale parameters of the input vectors $\boldsymbol{\mathsf{x}}$ which are essentially the columns of the $\boldsymbol{\mathsf{X}}$ matrix. In the description of our analytical transit model, we use the quadratic limb-darkening law\citeSupp{Kopal1950} to describe the centre-to-limb intensity variations across the stellar disk. This selection was made through a comparison of various alternative
laws, using the $\Delta$BIC formalism\citeSupp{Schwarz1978}. Finally, we generally assume uninformative, flat priors for the searched parameters, represented as $\mathrm{P} \left(\theta,\phi\right)$ in equation (\ref{eq:log posterior}), with the following exceptions:

\begin{equation}
\ln P(\boldsymbol{\theta},\boldsymbol{\phi}) = \begin{cases}
	\ln \mathcal{N}(0.78884,10^{-10}) & \text{for } \phi \in [\text{P}],\\
    \ln \mathcal{N}(0,10^{-10}) & \text{for } \phi \in [e],\\
	-\infty & \text{if $\left(c_1 + c_2 > 1\right)$},\\
	\ln \mathcal{N}(\mu_i,9\sigma_i^2) & \text{for } \phi \in [c_1,c_2],\\
    \ln \Gamma(1,1) & \text{if } \theta \in [\zeta,\eta_\alpha],\\
    -\infty & \text{if } \sigma < 0
	\end{cases}
\end{equation}

\noindent What these prior assumptions essentially mean is that we fix the period, P$_0$, and eccentricity, \textit{e}, of the transit to previously determined values, and set prior Gaussian distributions upon the values of the two limb-darkening coefficients, $c_1$ and $c_2$, on the basis of theoretically calculated values from PHOENIX models\citeSupp{Claret2013}, with a width three times larger than the uncertainty in the theoretical model values for a greater flexibility in the search space. A gamma function, $\Gamma$, is chosen as the prior for the parameters of the kernel in order to encourage their values towards 0 if they are truly irrelevant in describing the correlated noise in the data\cite{Gibson2012}.

We initially obtain values for the free parameters by using the Nelder–Mead simplex algorithm\citeSupp{Nelder1965}, and then find the maximum posterior solution by optimizing the log posterior given in equation (\ref{eq:log posterior}). We obtain the posterior probability distribution functions by using the adaptive MCMC method, which explores the joint probability for our multivariate models. For each light curve analysed here, we ran four independent MCMC simulations of length 100,000 iterations each, and checked their mutual convergence using the Gelman–Rubin\citeSupp{Gelman2014} diagnostic. Example sets of samples drawn from such chains are given in Extended Data Fig. \ref{fig:correlations} for a broadband and a spectral light curve fit. We note that in addition to our analytical transit model, we also include a baseline function that is a quadratic polynomial of the parallactic angle throughout the observations that accounts for modulations introduced to the light curves by the rotation of the telescope, as well as the second-order, colour-dependent extinction of the stars. This last fact is the reason why we fit for the parameters of the baseline model of each spectrophotometric light curve, independently.

We infer best-fit parameter solutions from the analysis of those derived posterior distributions from all the MCMC simulations. Given that all fitted parameters have a Gaussian posterior distribution, our results for the mean, mode and median of each parameter are statistically identical, with the exception of those noise parameters with gamma distribution prior functions. In those cases, we quote the median of the distribution as the solution (Extended Data Table \ref{tab:broadband res}).

\section*{Activity analysis}
It has been shown that transit light curves of WASP-19b towards the blue edge of the visible domain suffered substantially from the presence of stellar spots\cite{Huitson2013}. As a consequence, transmission spectra could not be produced from HST observations performed with the G430L grism of the Space Telescope Imaging Spectrograph (STIS) owing to the planet crossing such spot anomalies,
whose temperature difference compared with the surrounding stellar photosphere is larger towards shorter wavelengths. Therefore, we assume that our observations could possibly be affected by stellar activity, although not to the same extent as the HST results, because our wavelength coverage from the blue data set starts at 0.44 $\upmu$m (as compared with 0.29 $\upmu$m from STIS). There are two avenues through which presence of stellar spots can influence the transmission spectrum, namely occulted and unocculted spots.

\noindent \textit{Occulted spots.} Visual inspection shows that there is marginal evidence for a spot-crossing event in the blue data set after mid-transit. To calculate the impact of this anomaly, we modelled the light curves from the blue data set with a single spot-crossing event included as part of the analytical model. This approach meant the addition of four fitted parameters to the light curves to describe the physical properties of the spot anomaly: two positional variables (parallel and perpendicular to the transit chord), angular size (R$_\bullet$) in units of stellar radius (R$_\star$), and the contrast ratio (r$_\bullet$), which is wavelength dependent. We first modelled the broadband light curve from the blue data set as before, with the inclusion of four additional free parameters for the spot. This transit model including a spot-crossing event is shown in Extended Data Fig. \ref{fig:activity}a, quoting the derived spot parameters and the relative planetary radius and the limb-darkening coefficients. We detect a large spot anomaly overlapping with the transit chord, with a spectral radiance value very close to that of the surrounding photosphere, that is, a contrast ratio of 95\%. In order to obtain the spot temperature, we modelled all of the spectrophotometric light curves of the blue data set with a spot model, only fitting for the contrast ratio, as this is the only wavelength-dependent aspect of the spot, the relation for which is given by a differential application of Planck's law:

\begin{equation}\label{eq:Planck}
r_{\bullet} \equiv \frac{B_{\bullet}(\lambda,T)}{B_{\star}(\lambda,T)} = \frac{e^{hc/\left(\lambda k_BT_{\star}\right)}-1}{e^{hc/\left(\lambda k_BT_{\bullet}\right)}-1}
\end{equation}  

\noindent where $B$ is the spectral radiance of the spot ($\bullet$) or the star ($\star$); h is Planck's constant; \textit{c} is the speed of light; $k_B$ is Boltzmann's constant; $\lambda$ is the wavelength at which the flux is measured; and $T$ is the temperature of the spot or stellar photosphere.

The stellar photospheric temperature is 5,568 $\pm$ 71 K, obtained through precise spectroscopic measurements\citeSupp{Torres2012}. From fitting the spectral channels for a spot model, we obtained the wavelength dependence of the contrast ratio (Extended Data Fig. \ref{fig:activity}b). This relation is governed by equation (\ref{eq:Planck}), and, through a least-squares minimization approach, we obtained a value of 5,530 $\pm$ 10 K for the spot temperature--38 K cooler than the surrounding photosphere. Furthermore, we compared the initial transmission spectrum in the blue data set to that obtained by modelling the light curves with a spot model. These two sets of results are highly consistent, without any change to the obtained slope towards the shorter wavelengths (Extended Data Fig. \ref{fig:activity}c). This consistency highlights the strength of the Gaussian process red noise model in accounting for any systematic deviation from a transit function. We note that the error bars for the set that include a spot-crossing component are underestimated, because here no systematic noise model has been included. This is done to fully capture the impact of spot crossing
on the differential light curves. We also note that there are degeneracies between spot latitude, angular size and contrast. Thus our estimation of the spot contrast should be used with caution.

\noindent \textit{Unocculted spots.} WASP-19 is an active late G-star with spot anomalies on its surface that were detected by numerous previous studies\citeSupp{Tregloan2013,Mancini2013,Mandell2013}. Consequently, we have to assume that, in addition to the occulted spot detected here, there might be spots on the stellar surface that are not occulted by the planetary transit chord. The presence of such unocculted spots can lead to an overestimation of the integrated flux from the stellar disk, and therefore to an overestimation of the planetary relative radius, which is a chromatic effect. To estimate the impact of such unocculted spots on the transmission spectrum in the blue end of the spectrum and therefore their contribution to the observed slope, we simulated transit light curves for WASP-19b over a rotating spotted star, using the SOAP-T tooll\citeSupp{Oshagh2013SOAP}. For information minimization, we combined possible spots into a single spot with a 20\% filling factor, on the basis of previous spot detections in transit light curves of this planet.

To estimate the maximum impact of this unocculted spot on the transit depth measurements, we selected the longitude and latitude of the combined spot to be at the centre of the stellar disk, because this set-up maximizes the rotational modulations introduced into the photometry. The spot contrast is calculated via Planck's law, with the temperature of the spot set to 200 K, 600 K and 1,000 K below the surrounding photospheric temperature of 5,568 K. We then simulated transit light curves for each of the wavelength bins in the blue data set, at each given spot temperature, adapting those limb-darkening coefficient values that were used as priors in modelling the spectral channels. These light curves were then modelled in a similar manner to the spectrophotometric time series of the observed data, and the inferred relative radii as a function of wavelength were compared with the observational results for each given spot temperature (Extended Data Fig. \ref{fig:activity}c). The observed slope can be explained partially by the possible presence of unocculted spots on the stellar disk, given large spot temperature differences of more than 1,000 K. A spot temperature difference of about 500 K has been reported for an occulted spot\citeSupp{Tregloan2013}; this finding, together with our much lower temperature measurement (about 40 K below the photosphere for the occulted spot), suggests that the observed slope towards the blue end of the spectrum cannot be explained simply by the presence of unocculted spots on the surface of the star. However, any atmospheric conclusions made from measurements of this slope have to be used with this fact in mind. A similar analysis has been performed to determine the impact of unocculted spots on the transmission spectroscopy of HD 189733b, ruling out such an impact as the cause of the observed transmission spectral signals\citeSupp{McCullough2014} (that is, confirming the atmospheric nature of these signals).

\section*{Retrieval analysis}
\textit{Atmospheric models}. We modelled WASP-19b's atmosphere at the day–night terminator via 100 axially symmetric layers, uniformly spaced in log-pressure between 10$^{-6}$ and 10$^2$ bar. We take $R_p$ to be 1.31 $R_J$, $R_\star$ to be 0.993 $R_\odot$,
and a surface gravity of \textit{g} = 14.3 m s$^{−2}$ (where $R_p$ is the planetary radius, given in units of Jupiter’s equatorial radius, $R_J$, and $R_\star$ is the radius of the host star, given in units of the Sun’s radius, $R_\odot$). We assume hydrostatic equilibrium, along with terminator-averaged temperature structure\citeSupp{Madhusudhan2009} and chemistry, with each species distributed uniformly with altitude. Clouds are parameterized by an opaque cloud deck, below which no electromagnetic radiation may pass. Hazes are included via a two-parameter power law\cite{LecavelierDesEtangs2008}, with two free parameters $a$ and $\Upsilon$: $\sigma(\lambda) =  a \sigma_{0} (\lambda / \lambda_{0})^{\Upsilon} $, where $\sigma(\lambda)$ is the wavelength-dependent haze cross-section, $\lambda_0$ is a reference wavelength (350 nm) and $\sigma_0$ is the H$_2$–Rayleigh scattering cross-section at the reference wavelength (5.31 $\times$ 10$^{-31}$ m$^2$). Our usage of the terms `cloud' and `haze' refers to the spectral features they cause, and makes no distinction as to the formation mechanisms and microphysics involved. Inhomogeneous cloud and haze distributions across the terminator are considered, parameterized by a cloud-coverage factor\cite{MacDonald2017}.

The chemical composition of the atmosphere is modelled as H$_2$/He dominated, with a fixed H$_2$/He ratio of 0.17. This background gas is considered to contain a mixture of trace molecular and atomic species, each with parameterized volume mixing ratios ranging between 10$^{-16}$ and 10$^{-1}$ of the total atmospheric content. We consider a wide range of chemical species that have prominent absorption features at visible wavelengths, namely: H2O, Na, K, TiO, VO, AlO, TiH, NaH, MgH, CrH, CaH, ScH and FeH. We obtained the molecular line list for H$_2$O from the HITEMP\citeSupp{Rothman2010} database, while those for the metal oxides and hydrides
originated from the ExoMol project\citeSupp{Tennyson2012}. For each molecular species, cross-sections were pre-computed line-by-line\citeSupp{Hedges2016} and binned to a resolution of 1 cm$^{-1}$ on a grid of temperatures and pressures spanning the range 10$^{-4}$ bar to 10$^2$ bar, and 300 K to 3,500 K. The Na and K cross-sections were based on semi-analytic Lorentzian
line profiles\citeSupp{Christiansen2010}. H$_2$-H$_2$ and H$_2$-He collision-induced absorption is included from the HITRAN database\citeSupp{Richard2012}.

We computed the transmission spectrum of a given model atmosphere by integrating the stellar intensity, exponentially attenuated by the slant optical depth, over successive annuli for both cloud-free and uniformly cloudy terminator regions\cite{MacDonald2017}. The two resulting spectra were then linearly superimposed, weighted by the terminator cloud-coverage factor. The stellar intensities during and outside transit were integrated over the solid angle subtended at the observer to obtain their respective fluxes, with the transit depth given by the fractional stellar flux difference induced as the planet transits its host star. For the visible wavelength spectrum of WASP-19b, we computed the transit depth at 2,000 wavelength points, spaced uniformly
between 0.4 $\upmu$m and 1.1 $\upmu$m. Model spectra were convolved with the instrument
point spread functions and integrated over grism response curves to produce
binned model points corresponding to the blue, green and red data sets.

\noindent \textit{Retrieval methodology}. We explored the range of atmospheric properties consistent with WASP19b's transmission spectrum by using the POSEIDON\cite{MacDonald2017} atmospheric retrieval algorithm. This algorithm generates millions of model transmission
spectra, mapping the high-dimensional parameter space via the MultiNest\citeSupp{Feroz2008,Feroz2009,Feroz2013} multimodal nested sampling algorithm, implemented by the python wrapper PyMultiNest\citeSupp{Buchner2014}. This allows statistical estimation of the underlying atmospheric parameters, in addition to establishing detection significances for model features (for example, chemical species, clouds or hazes) via nested Bayesian model
comparison\citeSupp{Trotta2008}.

Each model atmosphere was parameterized by a six-parameter pressure–temperature profile\citeSupp{Madhusudhan2009}, a four-parameter inhomogeneous cloud/haze prescription\cite{MacDonald2017}, the \emph{a priori} unknown pressure at $R_\mathrm{p}$, $P_{\mathrm{ref}}$, and independent parameters for the volume mixing ratios of a subset of chemical species from Na, K, TiO, VO, AlO, TiH, NaH, MgH, CrH, CaH, ScH and FeH. Given the high resolution of the observed spectrum, we also allowed for the possibility of a systematic relative wavelength shift between the model and data as a nuisance parameter (often used in browndwarf retrievals of similar resolution\citeSupp{Line2015,Burningham2017}). This resulted in a maximum parameter space dimensionality of 25. Priors for each parameter were taken as either uniform (for example, top-of-atmosphere temperature) or uniform-in the logarithm (for example, mixing ratios), depending on whether the prior range was less than or more than two orders of magnitude, respectively.

We initially assessed plausibility of each chemical species by computing a full 25-dimensional retrieval with 2,000 MultiNest live points. We observed flat posteriors (no evidence, beyond an upper bound) for all chemical species except Na, K, H2O, TiO and MgH. The terminator cloud fraction sloped sharply towards the upper edge of the prior (100\% cloud/haze coverage). With respect to this reference model, we ran two identical retrievals—one with the cloud deck removed
and one with the haze removed. Removing the haze incurred a substantial penalty to the Bayesian evidence ($\ln \mathcal{Z} = 987.2 \rightarrow 970.2$), corresponding to a Bayes factor of 2.0 $times$ 10$^7$ and equivalent to a 6.2$\upsigma$ detection of a uniform haze across the terminator (with respect to the full chemistry model). The Bayesian evidence was unchanged within the error bars (about 0.1) when the cloud deck was removed, so the data do not support the presence of a cloud deck. We found a relative wavelength shift of $-72.8^{+10.0}_{-11.3}$ \AA~between the data and the model ($\ln \mathcal{Z} = 987.0 \rightarrow 970.6$, compared with a model in which the shift is fixed at 0, equivalent to a 6.1$\upsigma$ detection of a wavelength shift).

A second round of retrievals was conducted to assess the potential presence of Na, K, H$_2$O, TiO and MgH. Given the strong evidence for uniform hazes from the first round, we fixed the cloud fraction to 100\% for these runs, leading to a dimensionality of 16 for the reference model. We conducted a nested Bayesian model comparison by removing individual chemical species from the reference model, running a new retrieval for each, and computing the change in the Bayesian evidence. We observed little change, or only a very slight increase, when we removed K and MgH, indicating that these species were not detected and that their presence is suggested only marginally by the data. However, we observed changes in $\ln \mathcal{Z} > 4$ when Na, H$_2$O or TiO was removed, suggesting that they are present.

We conducted a final round of nested retrievals by using a minimal reference model containing just Na, H$_2$O, TiO, haze and a cloud deck--corresponding to 14 free parameters. This simple reference model has a Bayesian evidence of $\ln \mathcal{Z} = 990.71$, a notable increase over the initial reference model, reaffirming that the additional model complexity incurred by the large collection of other metal oxides/hydrides is not justified in light of the data. The results of the Bayesian model comparison, minimum best-fit reduced chi-squares, and detection significances
for Na, TiO, H$_2$O and the terminator-spanning haze, with respect to this reference model, are given in Extended Data Table \ref{table:model_comparison}. The overall best-fitting transmission spectrum, drawn from the posterior of the reference model (shown in red in Fig. \ref{fig:transmission_spectrum}), has Na, H$_2$O and TiO volume mixing ratios of 6.4 p.p.m., 180 p.p.m. and 0.12 p.p.b., respectively; haze parameters $a = 470,000$ and $\Upsilon = -30$; a wavelength shift of $-73.6$ \AA; and a reference pressure of 0.93 bar.

It has been claimed that accurate H2O abundances cannot be extracted from HST WFC3 transmission spectra\citeSupp{Heng2017}, unless a reference radius and pressure are assumed. This degeneracy may be lifted by observations of cloud-free atmospheres at visible wavelength, as the H$_2$-Rayleigh slope fixes the reference pressure\citeSupp{Heng2017}. Here, we have demonstrated that constraints of less than 1 dex can be placed on individual molecular abundances when retrieving the visible wavelength transmission spectrum of a hazy atmosphere, despite making no \emph{a priori} assumptions as to the value of $P_{\mathrm{ref}}$. Our ability to accomplish this stems from the data points that are less than about 0.5 $\upmu$m, which sample the continuum slope owing to our parametric haze model. This slope, in turn, provides a normalization to the transmission spectrum and enables a loose constraint to be placed on $P_{\mathrm{ref}}$, allowing the degeneracy with the molecular abundances to be partially collapsed. We stress that observations at short wavelengths, away from spectral features, are essential for constraining the absolute abundances of individual species.

\par
\noindent \textbf{Data availability.} The data used in this work can be accessed at the ESO science archive (archive.es o.org), using identification numbers 60.A-9203(F) for the green
data set, and 96.C-0465(B,C) for the blue and red data. Reduced one-dimensional and two-dimensional frames are available from E.S. upon reasonable request. In addition, the broadband (Fig. 1) and spectrophotometric light curves (Extended Data Figs 3--5), as well as the transmission spectrum data and models (Fig. 2), are included as source data online.

\noindent \textbf{Code availability.}  The GP (GeePea) and the MCMC inference (Infer) modules
are written in python programming language and are freely available from github.com/nealegibson. The multi-modal nested sampling algorithm MultiNest and its python wrapper PyMultiNest are freely available from https://ccpforge.cse.rl.ac.uk/gf/project/multinest/ and https://github.com/JohannesBuch ner/PyMultiNest. The SOAP-T tool is also freely available at http://astro.up.pt/resources/soap-t/.

\vspace{0.5cm}

%\bibliographySupp{WASP19}
%\bibliographystyleSupp{naturemag}

\newpage
\begin{figure}
\centering
\includegraphics[width=0.5\textwidth]{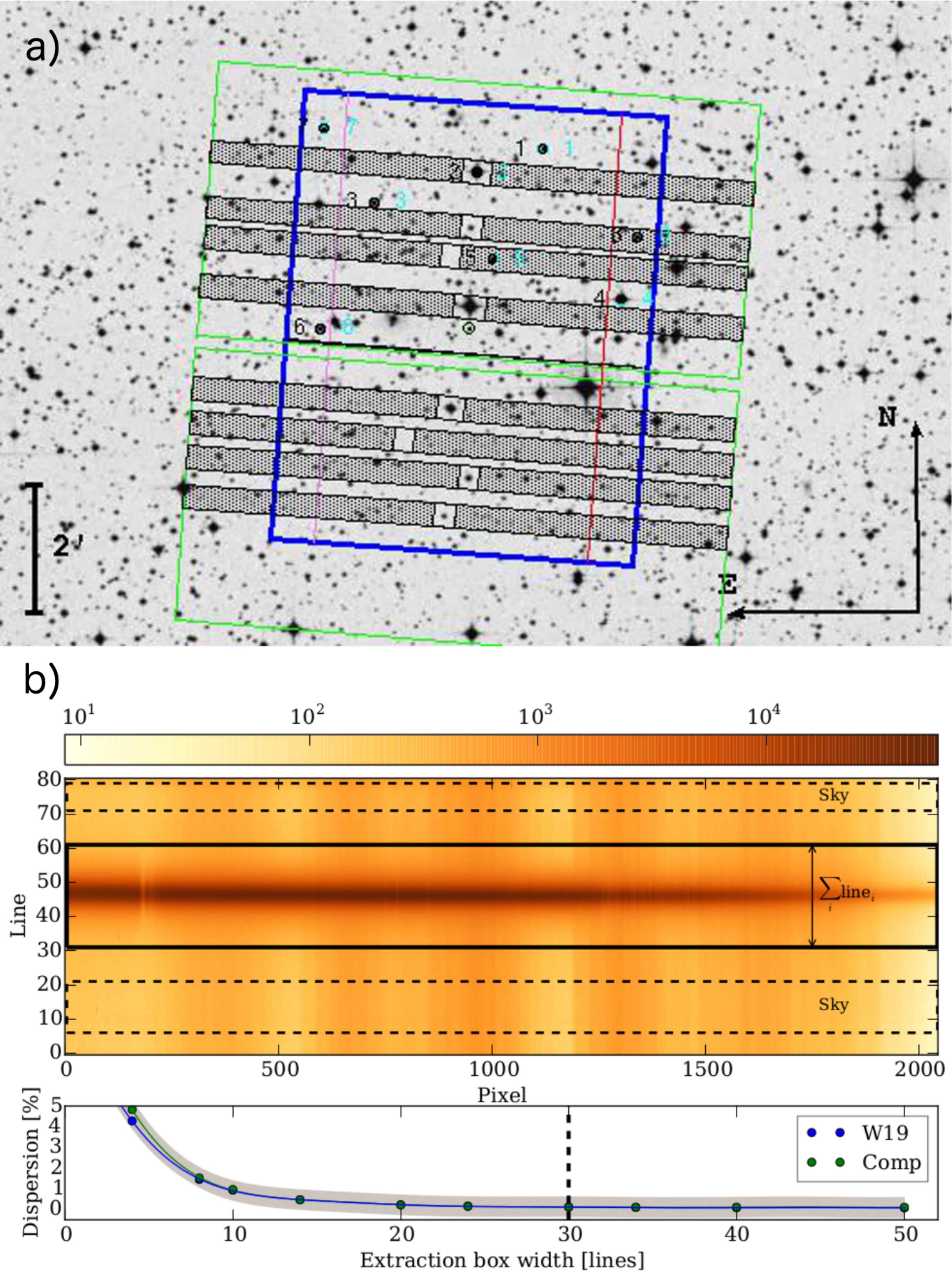}
\caption{\textbf{Observations and reduction.}  \textbf{a}, An example of a mask design used for MXU observations. The field of view of the FORS2 spectrograph is shown in blue, with the green lines indicating the two-chip detector mosaic. The grey shaded regions show the areas of the detectors used for recording the stellar spectra. In the instance shown, WASP-19 is the star in the upper-most slit. \textbf{b}, Top, an example of a two-dimensional spectrum, from the red data set, extracted from a frame taken with the FORS2 instrument. The final size of the extraction box and the regions used for sky subtraction are indicated. Bottom, the process of choosing the width of the extraction box, where the final value is shown as a dashed line and the grey shading represents the 1$\upsigma$ confidence limits. The exemplar frame used to produce these plots is selected at large seeing. W19, WASP-19; comp., comparison star.}
\label{fig:observations}
\end{figure}

\begin{figure}
\includegraphics[width=\textwidth]{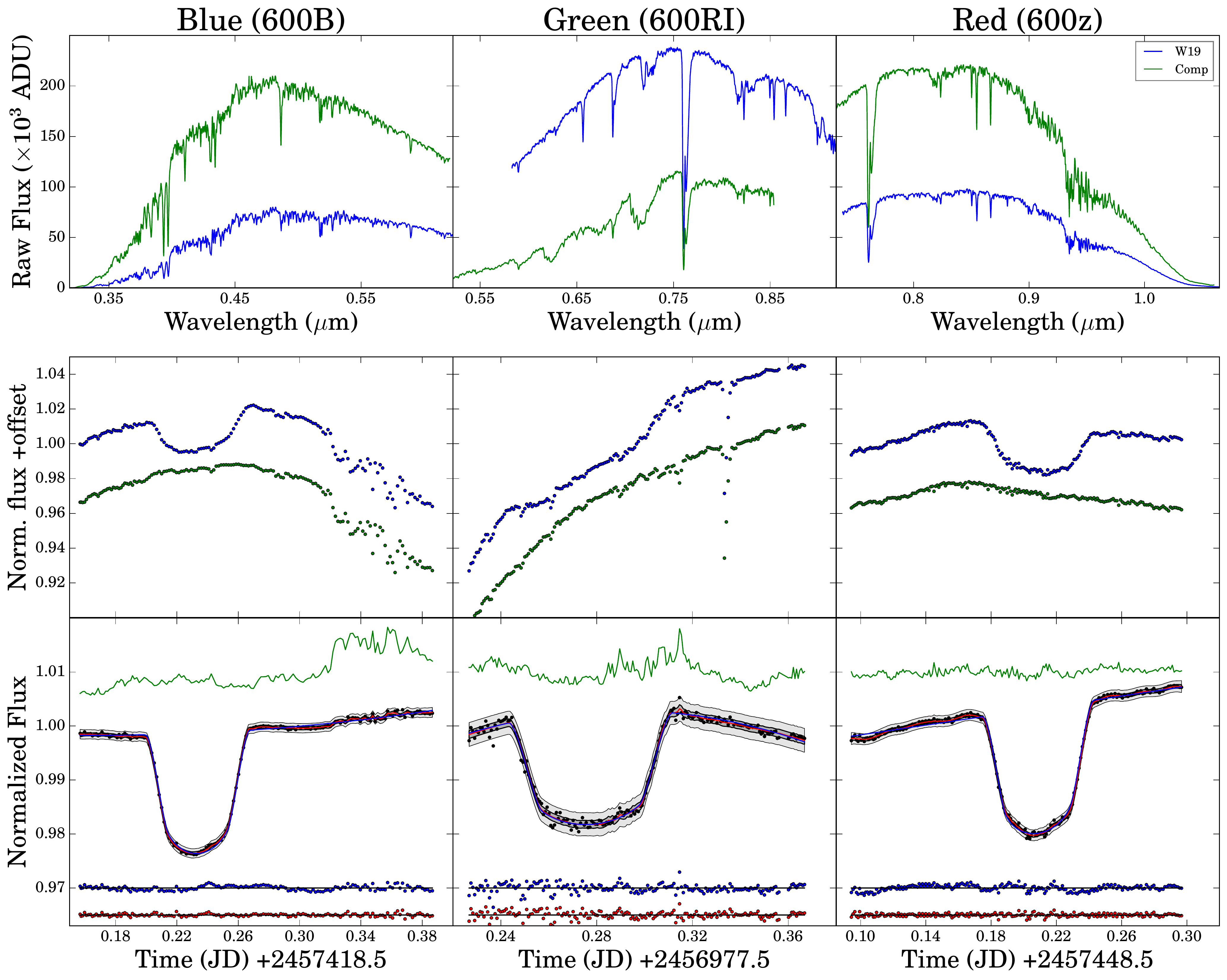}
\caption{\textbf{Spectra and light curves.}  Top row, an example set of spectra for the target (WASP-19) and the chosen comparison star, for each observing run (blue, green and red). Counts are given in analogue-to-digital units (ADUs), read directly from the sum of values of charge-coupled-device (CCD) pixels. Middle row, light curves for the target and comparison stars for each data set, obtained through broadband integration of the series of spectra. Colours match those in the top row; values are normalized to the mean of the out-of-transit fluxes and shifted
for clarity. The transit imprint from WASP-19b is clearly evident even in these raw light curves. Bottom row, differential broadband light curves (black data point) obtained simply by dividing the two light curves in the middle row. We also show our fitted transit model (blue curve) and the
Gaussian process systematic noise model (red curve) with its 1$\upsigma$ (dark grey shading) and 3$\upsigma$ (light grey shading) confidence levels. The points below are the residuals of the two models, where the colours correspond to the fit that they represent. The green line shows the flux variations resulting from changes in seeing conditions, used as an input for our Gaussian process model. JD, Julian day.}
\label{fig:LCs}
\end{figure}

\begin{figure}
\includegraphics[width=\linewidth]{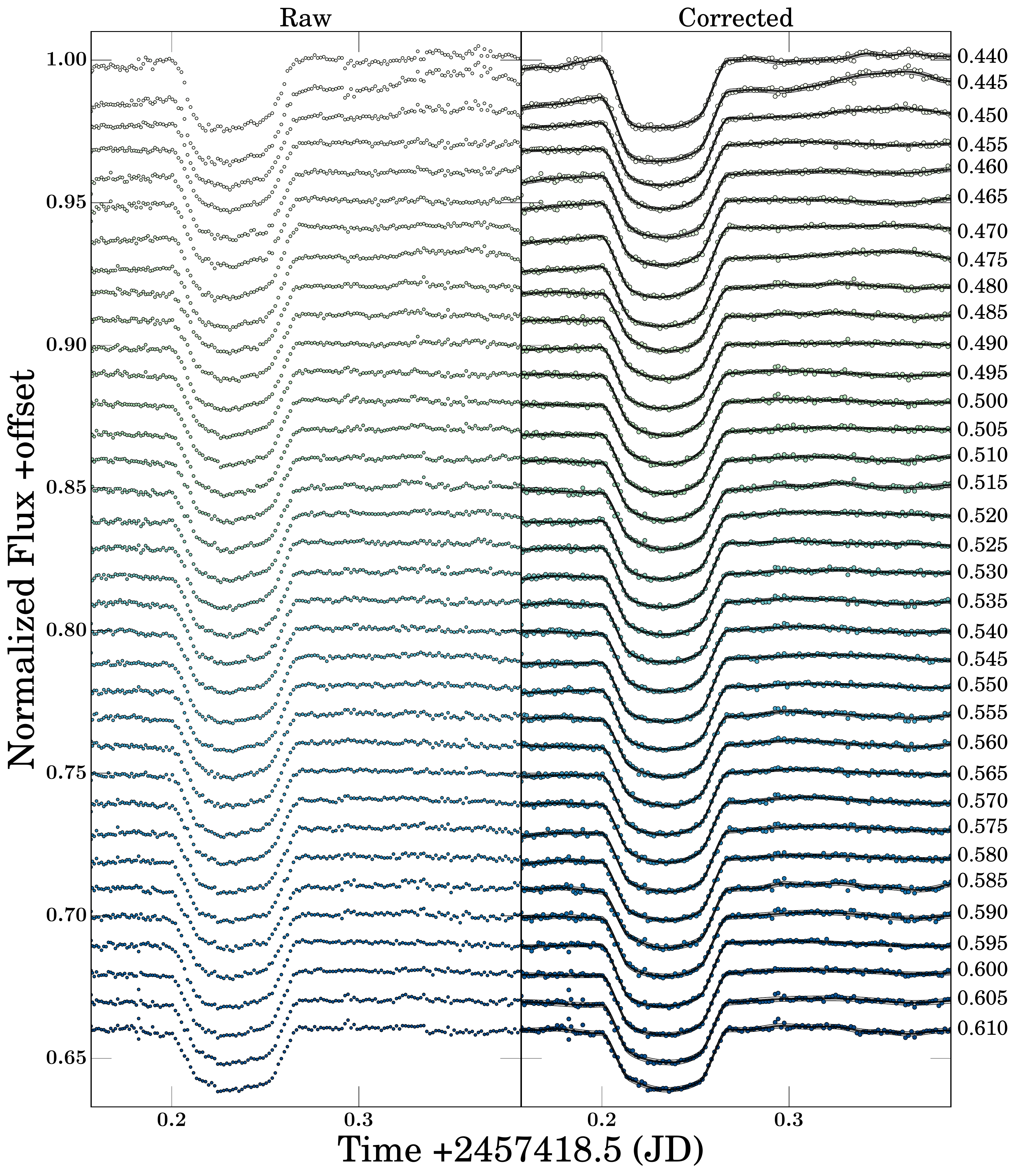}
\caption{\textbf{Spectrophotometric light curves - Blue.}  Left, raw light curves produced from each of the narrow-band channels in the blue data set. Right, those light curves that have been
corrected for the common-mode systematics. Our best-fit Gaussian process systematic noise models are shown as solid black lines, with the centre of the integration bin for each light curve given to the right of it in micrometres. All light curves have been shifted vertically for clarity.}
\label{fig:spec LCs - Blue}
\end{figure}

\begin{figure}
\includegraphics[width=\linewidth]{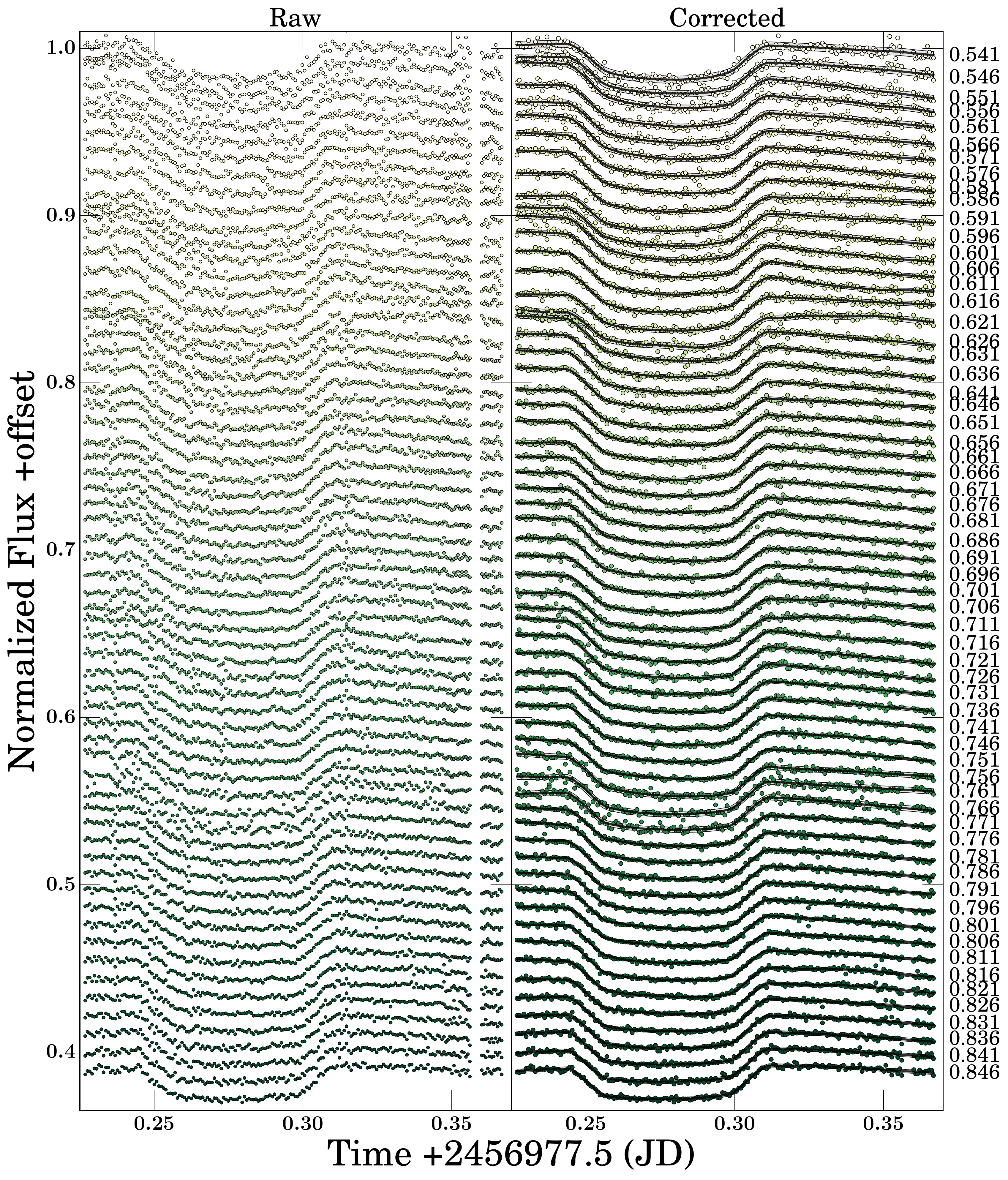}
\caption{\textbf{Spectrophotometric light curves - Green.}  As for Extended Data Fig. \ref{fig:spec LCs - Blue} but for the green data set.}
\label{fig:spec LCs - Green}
\end{figure}

\begin{figure}
\includegraphics[width=\linewidth]{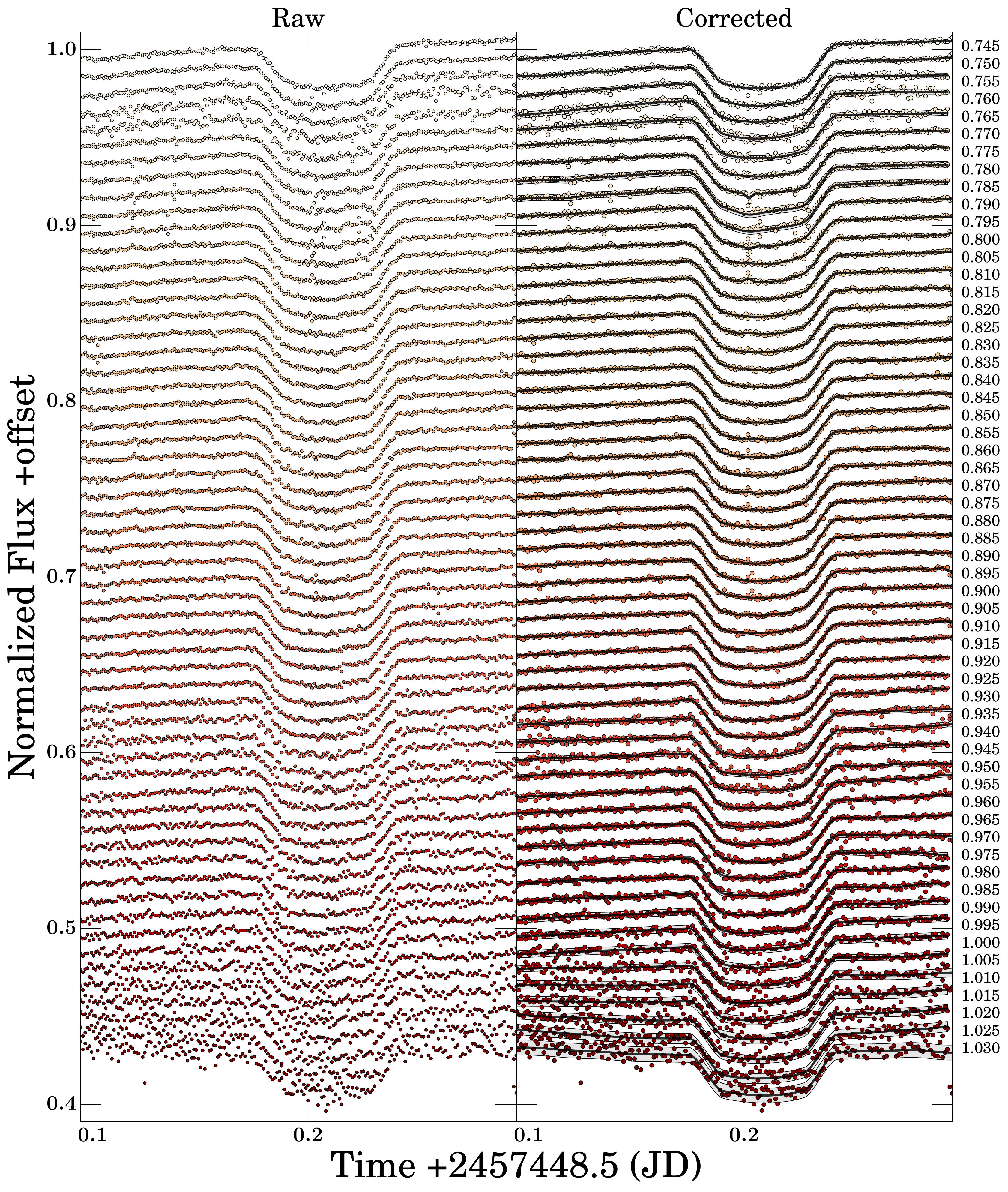}
\caption{\textbf{Spectrophotometric light curves - Red.}  As for Extended Data Figs \ref{fig:spec LCs - Blue} and \ref{fig:spec LCs - Green} but for the red data set.}
\label{fig:spec LCs - Red}
\end{figure}

\begin{figure}
\includegraphics[width=\linewidth]{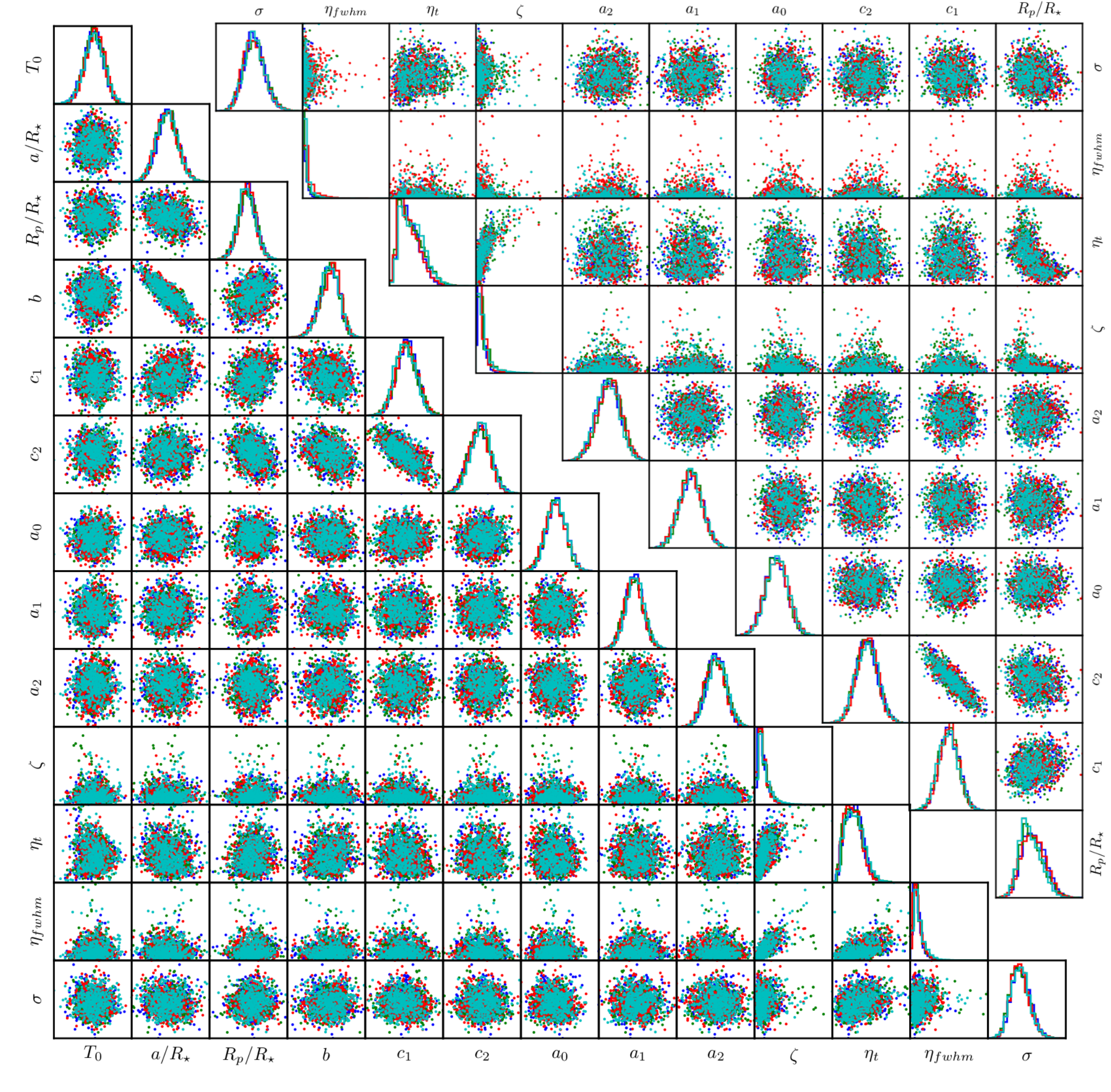}
\caption{\textbf{Correlations.} Random samples drawn from the four MCMC simulations, for all the fitted parameters (see Methods for definitions), in modelling a broadband light curve (lower-left triangle) and a spectroscopic light curve (upper-right triangle). Both examples are from the blue data set. Mutual convergence of all independent chains is evident, as are the well documented degeneracies between the impact parameter (b) and the scaled semi-major axis ($a / R_\star$), and between the two coefficients of the limb-darkening law ($c_1$ and $c_2$). $\eta_{\mathrm{fwhm}}$ is the Gaussian process inverse length scale for `seeing'.}
\label{fig:correlations}
\end{figure}

\begin{figure}
\centering
 \includegraphics[width=0.46\textwidth]{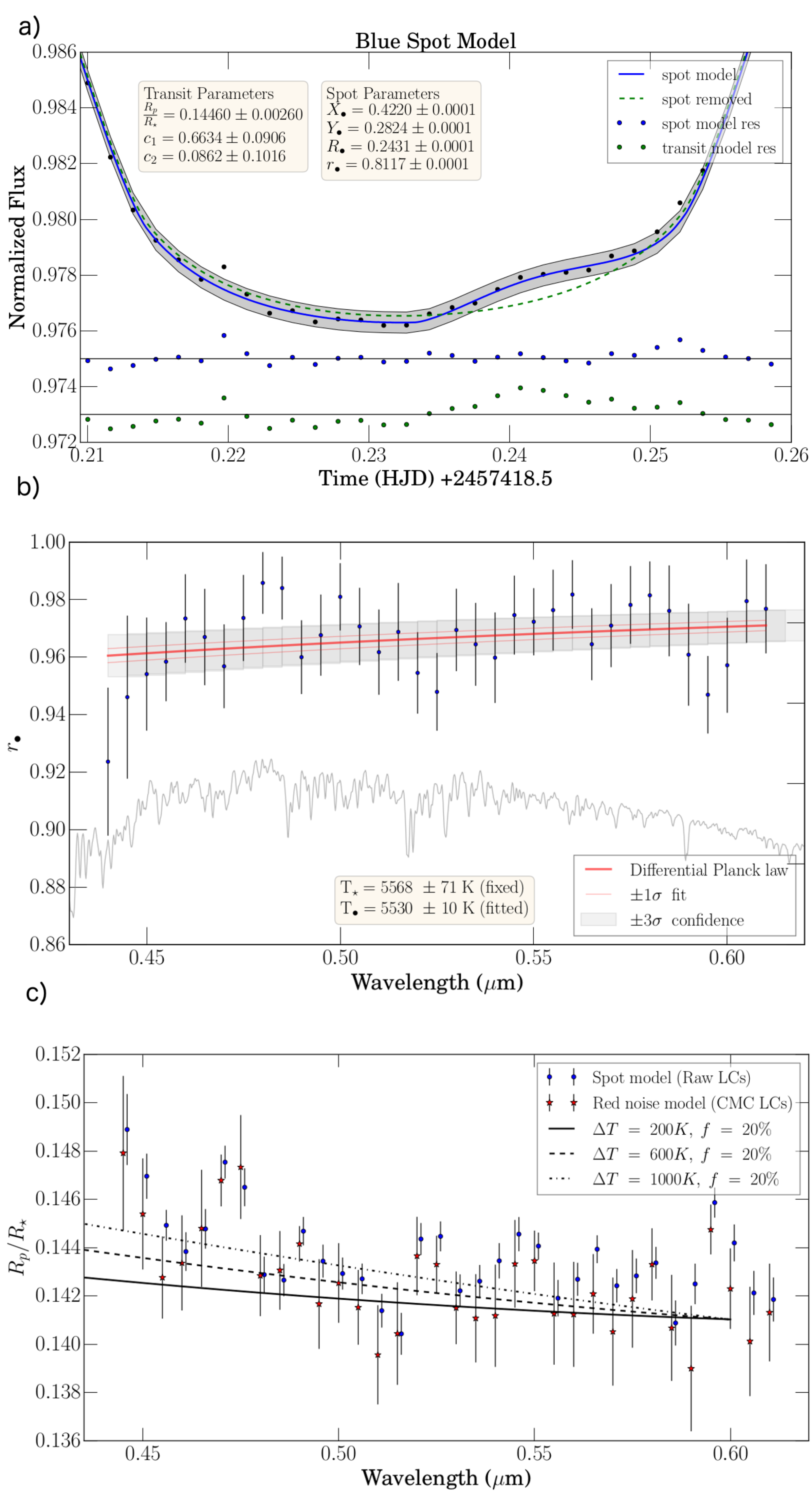}
 \caption{\textbf{Stellar activity impact.}  \textbf{a}, Broadband light curve from the blue data set, modelled using an analytical model that includes a spot-crossing event by the planet. The new inferred planetary radius and the limb-darkening coefficients (the `transit parameters') are shown, for which the offset to our previous results is substantially lower than the derived precision. This is because our systematic model accounts well for this anomaly. The inferred spot parameters are also shown. HJD, heliocentric Julian day; res, residuals. \textbf{b}, Dependence of spot contrast ratio on the observation wavelength, from which the spot temperature is determined using Planck's law. The 1$\upsigma$ error bars are derived from a joint analysis of posterior probability distributions of the relative planetary radius, from the MCMC simulations. For reference, a spectrum of WASP-19 is plotted in light grey. The prior stellar photospheric
temperature ($T_\star$) and the fitted spot temperature ($T_bullet$) are also given. \textbf{c}, Comparison of transmission spectra in the blue data set, from red noise analysis and spot modelling; 1σ error bars were derived as above. The spot-analysis results (blue points) have been shifted by +0.01 $\upmu$m to better distinguish between the two sets of results. Wavelength-dependent radius variations induced by the presence of spots with 20\% filling factor (f) and
temperature differences of 200 K, 600 K or 1,000 K are also plotted. LC, light curve; CMC, common mode corrected.}
 \label{fig:activity}
\end{figure}

\pagebreak
\thispagestyle{empty}
\linespread{1}
\newcommand{\ra}[1]{\renewcommand{\arraystretch}{#1}}
\begin{sidewaystable}
\ra{1.3}
\caption[]{\textbf{Observational information on the three sets of observational campaigns}}
%\begin{tabular*}{l c c c c c c c}
\begin{tabular*}{\columnwidth}{l@{\extracolsep{\fill}} l c c c c c c c@{}}\toprule
Dataset & Date & Airmass & Median & Grism & $\lambda$ & Readout & Exposures\\
 & & & seeing ($^{\prime\prime}$) & 600 & ($\upmu$m) & mode & (in-transit)\\
\midrule
Blue & 30-01-16 & 1.20$\rightarrow$1.07$\rightarrow$1.42 & 1.49 & B & 0.330-0.620 & 200kHz & 150 (43)\\
Green & 15-11-14 &2.57$\rightarrow$1.19 & 1.13 & RI & 0.536-0.853 & 100kHz & 180 (88)\\
Red & 29-02-16 & 1.15$\rightarrow$1.07$\rightarrow$1.47& 1.18 & z & 0.740\textbf{•}-1.051 & 200kHz & 212 (86)\\
\hline
\end{tabular*}
\label{tab:obs}
\end{sidewaystable}

\thispagestyle{empty}
\begin{sidewaystable}
\caption{\textbf{Transit parameters from broadband analysis of all three data sets.}}
\centering
\begin{tabular}{l c c c}
\hline
Parameter (inferred)  & Blue & Green & Red\\
 & 4,865 (2,530) \AA & 6,941.5 (3,172) \AA & 8,951.5 (3,111) \AA \\
\hline \hline
Mid-Transit, T$_c$ (JD) +2,400,000 & $ 57,418.73368\pm0.00010$ & $56,977.77754\pm0.00019$ & $57,448.70885\pm0.00020$ \\
T$_c$, Barycentric corrected (BJD$_{\text{TDB}}$) & 57,418.73688 & 56,977.77653 & 57,448.71294 \\
Period, P (days) [fixed] & \multicolumn{3}{c}{0.78884} \\
Eccentricity, $e$ [fixed] & \multicolumn{3}{c}{0.0} \\
Scaled semi-major axis, $a/R_{\star}$ & \multicolumn{3}{c}{$3.5875\pm0.0574$}\\
Impact parameter, $b$ & \multicolumn{3}{c}{$0.6525\pm0.0233$}\\
Relative planetary radius, $R_p/R_{\star}$ & $0.14440\pm0.00151$ & $ 0.14366\pm0.00181 $ & $ 0.14060\pm0.00148$ \\
Linear LD coefficient, $c_1$ & $ 0.6709\pm0.0616 $ & $ 0.5045\pm0.1037$ & $0.3152\pm0.1195$\\
Quadratic LD coefficient, $c_2$ & $ 0.0599\pm0.0866 $ & $ 0.1361\pm0.0934$ & $0.1673\pm0.1074$\\
GP output scale, $\zeta$ (ppm) & $748^{+418}_{~~~-220}$ & $650^{+1,723}_{~~~-218}$ & $644^{+308}_{~~~-181}$\\
GP inverse length scale for \textit{t}, $\eta_t$ & $0.028^{+0.010}_{~~~-0.009}$ & $0.027^{+0.093}_{~~~-0.015}$ & $0.008^{+0.002}_{~~~-0.001}$\\
GP inverse length scale for \textit{seeing}, $\eta_{\mathrm{fwhm}}$ & $0.006^{+0.004}_{~~~-0.002}$ & $0.010^{+0.018}_{~~~-0.003}$ & $0.006^{+0.013}_{~~~-0.002}$\\
Poisson noise, $\sigma_w$ (ppm) & $276\pm19$ & $729\pm71$ & $318\pm20$ \\
\hline
Parameter (derived) & & & \\
\hline \hline
Semi-major axis, $a$ (AU) & \multicolumn{3}{c}{0.01651$\pm0.00064$} \\
Orbital inclination, $i$ ($^{\circ}$) &  \multicolumn{3}{c}{$79.52^{+0.54}_{~~~-0.56}$}\\
Planet radius, $R_p$ ($R_{\mathrm{jup}}$) &$1.3907\pm0.0426$& $1.3836\pm0.0454$&$1.3541\pm0.0416$\\
\hline
\multicolumn{4}{l}{For each data set, the band centre and (in parentheses) the bandwidths are given at the top of the column. The data that are} \\
\multicolumn{4}{l}{shown only in the `green' column are common to all columns. GP, Gaussian process; LD, limb darkening; $R_{\mathrm{jup}}$, radius of}\\
\multicolumn{4}{l}{Jupiter.}
\end{tabular}
\label{tab:broadband res}
\end{sidewaystable}

\begin{table}
\ra{1.3}
\caption[]{\textbf{Bayesian model comparison detections of WASP-19b's terminator chemistry and cloud properties}}
\begin{tabular*}{\columnwidth}{l@{\extracolsep{\fill}} cccccl@{}}\toprule
Model & \multicolumn{1}{p{2cm}}{\centering \hspace{-0.5cm} Evidence \\ \centering $ \hspace{-0.3cm} \mathrm{ln}\left(\mathcal{Z}_{i}\right)$}  & \multicolumn{1}{p{2cm}}{\centering Best-fit \\ \centering $ \chi_{r, \mathrm{min}}^{2}$} & \multicolumn{1}{p{3cm}}{\centering \hspace{-0.3cm} Bayes Factor \\ \centering $ \hspace{-0.2cm} \mathcal{B}_{0i}$}& \multicolumn{1}{p{2cm}}{\centering \hspace{-0.4cm} Detection \\ \centering \hspace{-0.2cm} of Ref.}\\ \midrule
Reference & 990.71  &  1.63  & Ref. & Ref.\\
No TiO &  963.17  & 2.06 & 9.1 $\times$ 10$^{11}$ & \hspace{-0.5em} 7.7$\upsigma$ \\
No H$_2$O & 961.83 & 2.05 & 3.5 $\times$ 10$^{12}$ & 7.9$\upsigma$ \\
No Na & 986.36 & 1.65 & 77 & 3.4$\upsigma$ \\
No Haze & 965.33 & 1.95 & 1.0 $\times$ 10$^{11}$ & 7.4$\upsigma$ \\
\bottomrule
\vspace{0.1pt}
\end{tabular*}
The `reference' (ref.) model includes opacity resulting from H$_2$, He, Na, H$_2$O and TiO, along with a parameterized cloud and haze prescription. $\mathcal{Z}_i$ is the Bayesian evidence of the ith model; $\mathcal{B}_{0i}$ is the Bayes factor corresponding to the preference for the reference model over model $i$; and $\chi^2_{r, min}$ is the minimum reduced chi-squared value for the best-fitting spectrum within each model. An n$\upsigma$ detection (where n $\geq$ 3) indicates the degree of preference for the reference model over the alternative model.
\label{table:model_comparison}
\end{table}

\end{methods}
\end{document}